\documentclass[prd,showpacs]{revtex4}
\begin{document}

\newcommand{\re}{\mathop{\mathrm{Re}}}

\newcommand{\be}{\begin{equation}}
\newcommand{\ee}{\end{equation}}
\newcommand{\bea}{\begin{eqnarray}}
\newcommand{\eea}{\end{eqnarray}}

\title{On Energy of the Friedman Universes in Conformally Flat
Coordinates}

\author{Janusz Garecki}
\email{garecki@wmf.univ.szczecin.pl}
\affiliation{\it Institute of Physics, University of Szczecin, Wielkopolska 15,
          70-451 Szczecin, Poland}
\date{\today}

\input epsf

\begin{abstract}
Recently many authors have calculated energy of
the Friedman universes by using coordinate-dependent double index energy-momentum
complexes in Cartesian comoving coordinates $(t,x,y,z)$ and
concluded that the flat and closed Friedman universes are
energy-free. We show in this paper by using  Einstein canonical energy-momentum complex
and by doing calculations in conformally flat coordinates
that such conclusion is incorrect.
The results obtained in this paper are compatible with the results
of the our previous paper \cite{Gar07} where we have used coordinate-independent
averaged energy-momentum tensors to analyze Friedman universes.
\end{abstract}


\maketitle

\section{Introduction}

A spacetime is {\it conformally flat} if there exist coordinates $(\tau,x,y,z)$
in which the line element $ds^2$ reads
\begin{eqnarray}
ds^2 &=& \Omega^2(\tau,x,y,z)\bigl(d\tau^2 - dx^2-dy^2
-dz^2\bigr)\nonumber\\
&=& \Omega^2(\tau,x,y,z)\eta_{ik}dx^idx^k,
\end{eqnarray}
where $\eta_{ik}$ means the Minkowskian metric, i.e., $\eta_{ik} =
diag(1,-1,-1,-1)$ \footnote{We prefer signature$+---$and we will
use geometrized units in which $G=c=1$.}.
$\Omega(\tau,x,y,z)$ is a sufficiently smooth and positive-definite
function called {\it conformal factor}.

We will call the coordinates $(\tau,x,y,z)$ {\it the conformally
flat or conformally inertial} coordinates.

The conformally flat coordinates are determined up to
15-parameters group of the {\it conformal transformations}. This 15-parameters
Lie group contains, as a subgroup, the 10-parameters Poincare'
group \cite{Ing80, Nak03}.

It is obvious that the conformally flat coordinates are
{\it geometrically and physically distinguished} like inertial
coordinates $(t,x,y,z)$ in a Minkowskian spacetime \footnote{For example,
they determine the same causal structure underlying spacetime as
an inertial coordinates $(t,x,y,z)$ in a Minkowskian spacetime.}.

The necessary and sufficient   condition iff a four (or more)
dimensional spacetime could be conformally flat is vanishing of
its {\it Weyl conformal curvature tensor} \cite{Step03}.
Physically, the Weyl tensor describes source-free, i.e.,
independent of matter, gravitational field.

If a spacetime is neither {\it flat} nor {\it asymptotically flat} (at spatial
or at null infinity) but it is only {\it conformally flat}, then one should
prefer conformally flat coordinates to analyse physical properties
of the spacetime. Especially, one should prefer the conformally
flat coordinates in order to analyse energy and momentum  of such
spacetime by using {\it coordinate-dependent}, double index
energy-momentum complexes, matter and gravitation.

In this context we would like to remark that already in the case of a {\it
Minkowskian} spacetime the energy-momentum complexes can be reasonably used only
in an ``affine''  coordinates in which the metric components are
constant, eg., in an inertial (= Lorentzian) coordinates $(t,x,y,z)$
in which the line element $ds^2$ reads
\begin{equation}
ds^2 = dt^2 - dx^2 - dy^2 - dz^2.
\end{equation}

On the other hand, in an {\it asymptotically flat} spacetime one can
reasonably use these complexes only in an asymptotically flat (=
asymptotically inertial or asymptotically Lorentzian) coordinates.
So, in the case of a {\it conformally flat spacetime} one should use
the energy-momentum complexes in the conformally flat coordinates,
i.e., in the {\it conformally inertial coordinates}.

It is commonly known that the Friedman universes are conformally
flat \cite{Th78, Ligh75, Mas07}. So, it is natural to analyse of their
energetic content in conformally flat coordinates $(\tau,x,y,z)$.

Recently many authors have calculated energy of the
Friedman (and also more general, only spatially homogeneous)
universes \cite{one} mainly by using coordinate-dependent double index energy-momentum
complexes. These authors have performed their calculations not in
the conformally flat coordinates $(\tau,x,y,z)$ but in the
so-called {\it Cartesian comoving coordinates} $(t,x,y,z)$ in
which the line element $ds^2$ of the Friedman universes has the
form
\begin{equation}
ds^2 = dt^2 - {a^2(t)\bigl(dx^2 + dy^2 + dz^2\bigr)\over\bigl[1
+{k(x^2+y^2+z^2)\over 4}\bigr]^2},
\end{equation}
where $a = a(t)$ is the {\it scale factor}, and $k = 0,^+_-1$
means the {\it normalized curvature} of the slices $t = const$. $t$
denotes the universal time parameter called {\it cosmic time}.

In the Cartesian comoving coordinates $(t,x,y,z)$ only spatial
part of the full metric is conformally flat.

The above mentioned authors have concluded that the closed Friedman
universes have zero net global energy , and that the flat
Friedman universes are energy free, locally and globally
\footnote{I must say that in my old papers I also followed this conclusion.
Now I think that it was incorrect.}.
For an open Friedman universes one gets divergent global results in
the Cartesian comoving coordinates $(t,x,y,z)$.

In other comoving coordinates the results are dramatically
different (see, e.g., \cite{Gar07}).

Of course, the problem of the global quantities of the Friedman
and more general, only spatially homogeneous, universes {\it is
not well-posed} physical problem because {\it one cannot measure}
the global energy and momentum in the case. The global energy and
momentum, and global angular momentum also, have physical meaning
only in the case of an asymptotically flat spacetime (at spatial
or at null infinities) where these global quantities can be
measured. So, the calculations of the global energy and momentum,
and global angular momentum also, of an universe can have only
{\it some mathematical sense}.

In the case of an universe a physical sense can have only {\it
local quantities}, eg., energy density and its flux  and {\it global
quantities of an isolated part of the universe}, e.g., global energy of
the Solar System. If we use a coordinate-dependent double index energy-momentum complex, then
all these quantities should be calculated in {\it a
privileged coordinates}, e.g., in the case of a Friedman universe
one should use with this aim the geometrically and physically favorized {\it
conformally flat coordinates} $(\tau,x,y,z)$.

We would like to emphasize that the global result $E = 0$ obtained in
the Cartesian comoving coordinates $(t,x,y,z)$ for a closed Friedman universe is obtained
iff we admit the limiting process $r\longrightarrow\infty$
during integration over slice $t = const$, where $r=\sqrt{x^2 + y^2 + z^2}$
is the radial coordinate. But if $r\longrightarrow\infty $, then  the
spatial conformal factor ${a^2(t)\over(1 +{r^2\over 4})^2}$ {\it goes to
zero} in the case giving a singularity.

Resuming, one can doubt in physical validity of the conclusion that the closed
and flat Friedman universes (and also more general, only spatially
homogeneous Kasner and Bianchi universes) are energy free.

In this context, we would like to remark that by using our
coordinate independent {\it averaged relative energy-momentum
tensors} \cite{Gar07}  or {\it superenergy tensors} \cite{Gar93}
one can do {\it mathematically correct and coordinate independent} local analysis of
the Friedman and more general universes. One can also formally
calculate, correctly from the mathematical point of view, the
global, integral quantities for such universes.

It is interesting that following this way one gets {\it positive-definite} energy values for
the all Friedman universes and also for Kasner and Bianchi type I
universes \footnote{More general spatially homogeneous universes
have not been considered yet.}. So, in our opinion, all these
universes {\it needn't be}  energetic nonentity.

In this paper we present the results of the analysis of the
energetic content of the Friedman universes in the {\it distinguished}
conformally flat coordinates $(\tau,x,y,z)$. These coordinates
are the most suitable to this goal if one uses an energy-momentum complex.
Our analysis will be done with the help of
the most important in general relativity  Einstein's
canonical double index energy-momentum complex
\begin{equation}
_E K_i^{~k} := \sqrt{\vert g\vert}\bigl(T_i^{~k} + _E
t_i^{~k}\bigr) = _F {U_i^{~[kl]}}_{,l},
\end{equation}
where $_F{U_i^{~[kl]}} = (-)_F{U_i^{~[lk]}}$ mean Freud's
superpotentials, and $_E {t_i^{~k}}$ are the components of the
canonical Einstein's energy-momentum pseudotensor of the
gravitational field \cite{Tr62,Gold80}. $T_i^{~k}$ denote the
components of the symmetric energy-momentum tensor of matter.

As we will see, by using this energy-momentum complex in the
conformally flat coordinates $(\tau,x,y,z)$, {\it one cannot
assert} that the Friedman universes have zero net energy,
locally or globally.

The analogous result one can obtain by using any other
reasonable double index energy-momentum complex.

We hope that this paper and the our previous paper \cite{Gar07}
convincingly show that the Friedman universes  {\it are not energetic
nonentity}, neither locally nor globally.

Finishing this Section we would like to emphasize an important
superiority of the conformally flat coordinates $(\tau,x,y,z)$
over the Cartesian comoving coordinates $(t,x,y,z)$. Namely,
solving the energy-momentum problem of the Friedman universes in
Cartesian comoving coordinates $(t,x,y,z)$ one uses {\it only the line
element} (3) {\it independently} of the Einstein equations and their
solutions. On the other hand, the results obtained in conformally
flat coordinates $(\tau,x,y,z)$ {\it explicitly depend} not
only on the Friedman-Lemaitre-Robertson-Walker line element $ds^2$ but also on
{\it the solutions} of the Einstein equations.

In order to establish our attention we will consider in this
paper only dust Friedman universes.

The paper is organized as follows. In Section 2 we give dust Friedman
universes in conformally flat coordinates $(\tau,x,y,z)$, and
in Section 3 we will analyse the energy and its flux for dust
Friedman universes in these coordinates. Our analysis will be
performed with the help of the Einstein canonical
energy-momentum complex. Finally, in Section 4 we give our
conclusion.
\section{Dust Friedman universes in the conformally flat
coordinates $(\tau,x,y,z)$}
\vspace{0.3cm}
\subsection{Closed dust Friedman universes ( $k=1$)}
\vspace{0.3cm}

Let us consider the Friedman-Lemaitre-Robertson-Walker ({\bf
FLRW}) like line element
\begin{equation}
ds^2 = a^2(\eta)\bigl\{d\eta^2 - d\chi^2 -sin^2\chi\bigl(d\theta^2
+ sin^2\theta d\varphi^2\bigr)\bigr\}
\end{equation}
with the following ranges of the coordinates
$(\eta,\chi,\theta.\varphi)$:
\begin{equation}
0<\chi<\pi,~0<\theta<\pi,~0<\varphi< 2\pi,~ \chi-\pi<\eta<\pi -
\chi.
\end{equation}

Physically, the coordinate $\eta$ is the {\it conformal time}, $\chi$ is
a {\it radial coordinate}, and $\theta,\varphi$ are ordinary {\it spherical
angular coordinates} (see, e.g., \cite{Land80}).

The bijective transformation
\begin{eqnarray}
\tau + r = \tan\bigl({\eta+\chi\over 2}\bigr),~~\tau - r =
\tan\bigl({\eta-\chi\over 2}\bigr),\nonumber\\
\theta^{\prime}=\theta,~~\varphi^{\prime} = \varphi,\nonumber\\
0<\chi<\pi,~\chi -\pi<\eta<\pi-\chi,~0<\theta<\pi,~0<\varphi<
2\pi,
\end{eqnarray}
with inverse
\begin{eqnarray}
\eta&=& \arctan(\tau + r) + \arctan(\tau-r),~(-)\infty<\tau +
r<\infty,\nonumber\\
\chi&=& \arctan(\tau + r) - \arctan(\tau -r),~(-)\infty<\tau
-r<\infty,~0<r<\infty,\nonumber\\
\theta&=&\theta^{\prime},~~\varphi =
\varphi^{\prime},~0<\theta^{\prime}<\pi,~0<\varphi^{\prime}< 2\pi,
\end{eqnarray}

map this spacetime onto conformally flat spacetime with the
following line element
\begin{eqnarray}
ds^2 = {4a^2(\tau,x,y,z)\over[1+(\tau+r)^2][1+(\tau-r)^2]}\eta_{ik}dx^idx^k\nonumber\\
&=:& \Omega^2(\tau,x,y,z)\eta_{ik}dx^idx^k,
\end{eqnarray}
where
\begin{eqnarray}
x&=& r\sin\theta\cos\varphi,~y = r\sin\theta\sin\varphi,~z =
r\cos\theta,\nonumber\\
r&=& \sqrt{x^2+y^2+z^2}.
\end{eqnarray}

This means that the transformation (7) covers the region of the
spacetime (5)-(6) with the conformally flat coordinates
$(\tau,x,y,z)$. One can call these coordinates {\it the
conformally inertial coordinates}.

If we omit the angular coordinates $(\theta,\varphi)$ then this
region will be the triangle
\begin{equation}
0<\chi<\pi, ~~\chi -\pi<\eta<\pi-\chi
\end{equation}
on the plane $(\eta,\chi)$.

Now let us consider a closed dust Friedman universe with the
following line element in the same coordinates $(\eta,\chi,\theta,\varphi)$
\begin{equation}
ds^2 = a^2(\eta)\bigl\{d\eta^2 - d\chi^2 -
\sin^2\chi\bigl(d\theta^2 + \sin^2\theta d\varphi^2\bigr)\bigr\},
\end{equation}
with
\begin{equation}
a = a_0\bigl(1 + \cos\eta\bigr), ~~t = a_0 \bigl(\eta +\pi +
\sin\eta\bigr),
\end{equation}
and with the following ranges of the coordinates $(\eta,\chi,\theta,\varphi)$
\begin{equation}
(-)\pi<\eta<\pi, ~0<\chi<\pi,~~0<\theta<\pi,~~0<\varphi< 2\pi.
\end{equation}

The coordinates $(\eta,\chi,\theta,\varphi)$ are comoving, i.e.,
the dust particles and so-called {\it fundamental observers} are
at rest in these coordinates.

$a(\eta)$   is the {\it scale factor}  and $t$ means the {\it
cosmic time}; $a_0 = {4\over 3} \pi\rho a^3 = const $ is the first
integral of the Friedman equations in the case.

If we omit the angular coordinates $(\theta,\varphi)$, then this
universe is a rectangle $(-)\pi<\eta<\pi,~0<\chi<\pi$ on the plane
of the variables $\eta,~\chi$.

Comparing this rectangle with the previous triangle one can easily
see that the conformally flat coordinates $(\tau,x,y,z)$ cover
only this half of the closed dust Friedman universe which is
determined by the following ranges of the coordinates $(\chi,\eta,\theta,\varphi)$
\begin{equation}
0<\chi<\pi,~\chi -\pi<\eta<\pi-\chi,~0<\theta<\pi,~0<\varphi< 2\pi.
\end{equation}

It is worth to emphasize that only one slice, $\eta=0$, of the closed dust Friedman universe
is {\it entirely covered} by the conformally flat coordinates $(\tau = 0,x,y,z)$.
Any other slice $\eta = \eta_0\not= 0$ is only {\it partially
covered} by these coordinates.

Applying an active point of view one can say that this distinguished slice $\eta = 0$
is mapped onto subspace
\begin{equation}
\tau = 0, ~
(-)\infty<x<\infty,~(-)\infty<y<\infty,~(-)\infty<z<\infty
\end{equation}
of the conformally flat spacetime $(\tau,x,y,z)$ which has the
line element (9) with
\begin{equation}
a(\tau,x,y,z)= a_0\bigl\{ 1 +\cos[\arctan(\tau + r) + \arctan(\tau
- r)]\bigr\}.
\end{equation}
The limiting values
\begin{equation}
x = ^+_-\infty,~y = ^+_-\infty,~z = ^+_-\infty
\end{equation}
{\it are not admissible} by the condition $\Omega(0,x,y,z)>0$.

It follows that in conformally flat coordinates it is possible to
calculate integrals only over the distinguished spatial slice
$\eta=0$ \footnote{$\eta=0$ corresponds to space $\tau = 0$ in the
conformally flat coordinates $(\tau,x,y,z)$ as it was already
mentioned before.}. This fact is very important, e.g., for
formal calculating global energy  and momentum of a dust closed
Friedman universe.

It is very interesting that in the conformally flat coordinates $(\tau,x,y,z)$
the initial singularity at $\eta = (-)\pi$   and the final
singularity at $\eta = \pi$ are removed to $\tau = (-)\infty$ and
to $\tau = \infty$ respectively, i.e., we {\it have no cosmological
singularity} in the case at a finite moment of the conformal time coordinate $\tau$.

Matter and comoving (= fundamental) observers are not at rest in
the conformally flat coordinates $(\tau,x,y,z)$. They both move
with the same 4-velocity
\begin{eqnarray}
u^0 &=& {1+\tau^2 +r^2\over 2a(\tau,x,y,z)},~u^1 =
{\sin\theta\cos\varphi\cdot \tau\cdot r\over
a(\tau,x,y,z)},\nonumber\\
u^2&=& {\sin\theta\cos\varphi\cdot\tau\cdot r\over
a(\tau,x,y,z)},~u^3 = {\cos\theta\cdot\tau\cdot r\over
a(\tau,x,y,z)},
\end{eqnarray}
where
\begin{eqnarray}
a &=& a_0\bigl\{ 1 + \cos[\arctan(\tau + r) + \arctan(\tau -
r)]\bigr\},\nonumber\\
\sin\theta&=& {\sqrt{x^2 + y^2}\over r}, ~\cos\theta = {z\over r},
~\cos\varphi = {x\over\sqrt{x^2+y^2}},\nonumber\\
\sin\varphi &=&{y\over\sqrt{x^2+y^2}},~r = \sqrt{x^2+y^2+z^2}.
\end{eqnarray}
Only fundamental observers which lie in the distinguished slice $\eta = 0$
remain also at rest in the conformally flat coordinates $(\tau,x,y,z)$
in the slice $\tau = 0$.

\vspace{0.3cm}
\subsection{Open dust Friedman universe ($k=-1)$}
\vspace{0.3cm}
Now, let us consider an open dust Friedman universe endowed with
the same comoving coordinates $(\eta,\chi,\theta,\varphi)$ as in
the closed case.

We have (see, e.g., \cite{Land80})
\begin{eqnarray}
ds^2 &=& a^2(\eta)\bigl\{d\eta^2 -d\chi^2 -
\sinh^2\chi\bigl(d\theta^2
+\sin^2\theta d\varphi^2\bigr)\bigr\},\nonumber\\
a&=& a_0(\cosh\eta-1),~~t =a_0(\sinh\eta-\eta),
\end{eqnarray}
where   $a_0 = {4\over 3}\pi\rho a^3 = const$, and
\begin{equation}
0<\eta<\infty,~0<\chi<\infty,~0<\theta<\pi,~0<\varphi< 2\pi.
\end{equation}
Then, the transformation
\begin{eqnarray}
r& =& {a_0\over 2} e^{\eta}\sinh\chi,~ \tau = {a_0\over
2}e^{\eta}\cosh\chi,~\tau>{a_0\over 2}, ~r>0,\nonumber\\
\theta^{\prime}&=&\theta,~\varphi^{\prime} = \varphi,
\end{eqnarray}
with inverse
\begin{eqnarray}
\eta &=& \ln\bigl({2\sqrt{\tau^2 - r^2}\over
a_0}\bigr),~\tau^2-r^2>{a_0^2\over
4},~\theta=\theta^{\prime},~\varphi=\varphi^{\prime},\nonumber\\
\tanh\chi &=& {r\over\tau}\longrightarrow \sinh\chi =
{\tau^2\over\tau^2-r^2},
\end{eqnarray}
brings the line element (21)-(22) to the conformally flat form
\begin{eqnarray}
ds^2 = \bigl(1 - {a_0\over
2\sqrt{\tau^2-r^2}}\bigr)^4\eta_{ik}dx^idx^k\nonumber\\
&=:& \Omega^2(\tau,x,y,z)\eta_{ik}dx^idx^k.
\end{eqnarray}
Here the {\it conformal factor} $\Omega = \bigl(1-{a_0\over
2\sqrt{\tau^2-r^2}}\bigr)^2$, and $\tau^2 -r^2 >{a_0\over 2}$. $r
=\sqrt{x^2+y^2+z^2}$, $x = r\sin\theta\cos\varphi,~y = r\sin\theta\sin\varphi,~z = r\cos\theta.$

From an active point of view the transformation (23) maps the open
dust Friedman universe (21)-(22) onto interior of the future
light cone $\tau^2 -x^2-y^2-z^2 = 0$ of a Minkowskian spacetime
which line element in an inertial coordinates  reads
\begin{equation}
ds^2 = \eta_{ik}dx^idx^k.
\end{equation}
Under this mapping a slice $0<\eta=\eta_0$ of the open dust
Friedman universe is mapped onto hyperboloid $\tau^2 - r^2 = B^2, ~B^2 := {a_0^2 e^{2\eta_0}\over
4}$ in the spacetime with the line element (26).

In the conformally flat coordinates $(\tau,x,y,z)$ the dust
matter filling the open Friedman universe and comoving
fundamental observers also {\it are not at rest}. Namely, they
have the following 4-velocity in these coordinates
\begin{equation}
u^0 = {\tau\over a}, ~u^1 = {x\over a},~u^2 = {y\over a},~u^3 =
{z\over a},
\end{equation}
where
\begin{equation}
a = a_0\bigl({\tau^2 -r^2 +a_0^2/4\over a_0\sqrt{\tau^2 - r^2}}
-1\bigr), ~r^2 = x^2+y^2+z^2.
\end{equation}
\vspace{0.3cm}
\subsection{Flat dust Friedman universes ($k=0$)}
\vspace{0.3cm}
Finally, let us consider a flat Friedman universe filled with dust
matter in the Cartesian comoving coordinates $(t,x,y,z)$.

We have (see, e.g., \cite{Land80})
\begin{equation}
ds^2 = dt^2 -a^2(t)\bigl(dx^2+dy^2+dz^2\bigr),
\end{equation}
where
\begin{equation}
a(t) = At^{2/3}, ~~A  = {4\over 3}\pi\rho a^3 = const>0, ~0<t<\infty.
\end{equation}
The parameter $t$ is the {\it cosmic time} and $a(t)$ denotes as usual the {\it
scale factor}.

In order to pass to the conformally flat coordinates $(\tau,x,y,z)$
it is sufficient in the case only to change the time coordinate $t$
onto {\it conformal time} $\tau$ following the scheme
\begin{equation}
d\tau = {dt\over a(t)}.
\end{equation}
From (30)-(31) it follows that
\begin{equation}
\tau = {3\over A}t^{1/3}\equiv t = {A^3\over 27}\tau^3,
\end{equation}
and
\begin{equation}
a(\tau) := a[t(\tau)] = {A^3\over 9}\tau^2, ~~0<\tau<\infty.
\end{equation}

Substituting into line element (29) $dt^2 = a^2(\tau)d\tau^2$ we
get
\begin{equation}
ds^2 = a^2(\tau)\bigl(d\tau^2 - dx^2-dy^2-dz^2\bigr),
\end{equation}
i.e., we get the line element (29)-(30) in the {\it conformally
flat form} with the {\it conformal factor}
\begin{equation}
\Omega = \Omega(\tau) = a(\tau) = {A^3\over 9}\tau^2,
~0<\tau<\infty.
\end{equation}

From geometrical point of view the flat dust Friedman universe  in
conformally flat coordinates $(\tau,x,y,z)$ is identical with the
upper half $(\tau>0)$ of the conformally flat spacetime which has the
following line element
\begin{equation}
ds^2 = a^2(\tau)\bigl(d\tau^2 - dx^2-dy^2-dz^2\bigr).
\end{equation}

It is interesting that in this case the conformally flat
coordinates $(\tau,x,y,z)$ are also {\it comoving coordinates},
like initial Cartesian coordinates $(t,x,y,z)$.

The 4-velocity of a particle of the dust which fills the flat
Friedman universe (identical with the 4-velocity of a fundamental
observer) reads in the conformally flat coordinates $(\tau,x,y,z)$
\begin{equation}
u^i = {\delta^i_0\over a(\tau)} \equiv u_i = a(\tau)\eta_{io}.
\end{equation}

It results  that the dust and the fundamental observers both {\it are at
rest} in these coordinates, like  as in the Cartesian comoving
coordinates $(t,x,y,z)$.
\vspace{0.3cm}
\section{Energy of the Friedman universes in the conformally flat
coordinates $(\tau,x,y,z)$}
\vspace{0.3cm}

In this Section we will consider the energetic content of the
Friedman universes in the physically and geometrically
distinguished {\it conformally flat coordinates} $(\tau,x,y,z)$.
We will use in our analysis the double index Einstein's canonical
energy-momentum complex, matter and gravitation,
\begin{equation}
_E K_i^{~k} := \sqrt{\vert g\vert}\bigl(T_i^{~k} + _E
t_i^{~k}\bigr).
\end{equation}
Here $T_i^{~k}$ are the components of the symmetric
energy-momentum tensor of matter and $_E t_i^{~k}$ mean the
components of the so-called {\it Einstein gravitational
energy-momentum pseudotensor} (see, e.g., \cite{Tr62, Gold80,
Land80}).

It is known that
\begin{equation}
\sqrt{\vert g\vert}\bigl(T_i^{~k} + _E
t_i^{~k}\bigr) = _F {U_i^{~[kl]}}_{,l},
\end{equation}
where $_F {U_i^{~[kl]}} = (-) _F {U_i^{~[lk]}}$ are {\it Freud's
superpotentials} which in a coordinate basis read
\begin{equation}
_F {U_i^{~[kl]}} = \alpha\bigl\{{g_{ia}\over\sqrt{\vert
g\vert}}\bigl[(-g)\bigl(g^{ka}g^{lb} - g^{la}
g^{kb}\bigr)\bigr]_{,b}\bigr\},~~\alpha = {1\over 16\pi},
\end{equation}
and that the equations (39) represents special form of the
Einstein equations (in mixed form and multiplied by $\sqrt{\vert
g\vert}$).

Owing antisymmetry  of the Freud superpotentials one can easily
obtain from (39) the following {\it local energy-momentum
conservation laws}, for matter and gravitation
\begin{equation}
_E {K_i^{~k}}_{,k}= 0.
\end{equation}

By using integral Stokes theorem one can obtain from(41)
reasonable {\it integral conservation laws} for a closed system in
an asymptotically flat coordinates.

Of course, one can consider in {\bf GR}  many other
energy-momentum complexes. But the Einstein expressions is the
best one of the all variety of the energy-momentum complexes (see, e.g., \cite{Gold80}).
In consequence, we confine in this paper, like in our previous papers, only to this
double index energy-momentum complex\footnote{But using of an other reasonable double
index energy-momentum complex will lead us to analogous results.}.

For a conformally flat spacetime with
\begin{eqnarray}
g_{ik}&=&\Omega^2\eta_{ik}\equiv g^{ik} = \Omega^{-2}
\eta^{ik},~\Omega = \Omega(\tau,x,y,z),\nonumber\\
\sqrt{\vert g\vert}&=& \Omega^4,
\end{eqnarray}
one obtains from (39)-(40)
\begin{equation}
_E K_i^{~k} = 4\alpha\bigl(\delta^k_i{}\eta^{lb} -
\delta^l_i{}\eta^{kb}\bigr)\bigl(\Omega_{,l}\Omega_{,b} + \Omega
\Omega_{,bl}\bigr).
\end{equation}
As a trivial conclusion we get from (43)
\begin{equation}
_E K_0^{~0} = 0
\end{equation}
iff $\Omega = \Omega(x^0)\equiv\Omega(\tau)$.

We have the situation of such a kind in the case of a flat
Friedman universe.

Note that in this case the component $_E K_0^{~0}$ has physical
meaning of the total ``energy density'', matter and gravitation,
for comoving observers which have 4-velocities $u^i = {\delta^i_0\over
a(\tau)}$.

In general, simple calculations performed by using  (43) and
formerly given forms of the conformal factor $\Omega(\tau,x,y,z)$ for the considered
dust Friedman universes lead us to the following results:
\begin{enumerate}
\item In the case of a flat, dust Friedman universe only the
components
\begin{equation}
_E K_1^{~1} =_E K_2^{~2} = _E K_3^{~3} = 4\alpha\bigl({\dot a}^2 +
a{\ddot a}\bigr)
\end{equation}
of the canonical energy-momentum complex $_E K_i^{~k}$ are
different from zero in the conformally flat coordinates
$(\tau,x,y,z)$.
Here ${\dot a} :={da\over d\tau}, ~{\ddot a}:= {d^2a\over
d\tau^2}$.
Thus, in the case, {\it not the all components} of the complex $_E K_i^{~k}$ {\it
are vanishing}.

In consequence, there exist observers with  4-velocities
\begin{eqnarray}
u^i&=& \bigl({1\over a\sqrt{1-v^2}},~{v_x\over
a\sqrt{1-v^2}},~{v_y\over a\sqrt{1-v^2}},~{v_z\over
a\sqrt{1-v^2}}\bigr),\nonumber\\
v_x&=& {dx\over d\tau},~v_y = {dy\over d\tau},~v_z = {dz\over
d\tau}, ~v^2 = v_x^2 + v_y^2 +v_z^2,
\end{eqnarray}
for which  the ``energy density'' $\epsilon:= _E K_i^{~k}u^iu_k$
and its flux (=Poynting's vector)
\begin{equation}
P^i = \bigl(\delta^i_k - u^iu_k\bigr)_E K_l^{~k}u^l
\end{equation}
are different from zero.

Namely, we have for such observers
\begin{eqnarray}
\epsilon&=& {(-)8\over 27}\alpha
A^6\tau^2{}{v^2\over(1-v^2)}<0,\nonumber\\
P^0&=& {4\alpha({\dot a}^2 + a{\ddot a})v^2\over
a(\tau)(1-v^2)^{3/2}}, ~P^{\beta} = {4\alpha({\dot a}^2 + a {\ddot
a})v^{\beta}\over a(\tau)(1-v^2)^{3/2}},
\end{eqnarray}
where
\begin{equation}
a(\tau) = {A^3\over 9}\tau^2>0,~{\dot a} = {2A^3\over 9}\tau>0,
~{\ddot a} = {2A^3\over 9}>0,~\beta = 1,2,3.
\end{equation}

The formal integral
\begin{equation}
E = \int\limits_{\tau = const}\epsilon dxdydz
\end{equation}
is divergent to minus infinity.

We would like to remark that the spatial velocity $v^2 = v_x^2 + v_y^2 + v_z^2$
of these observers can be {\it infinitesimally small}, i.e., these
observers can {\it infinitesimally} differ from comoving
observers.

Only for {\it comoving observers} which have their 4-velocity of
the form $u^i = {\delta^i_0\over a}$ we have
\begin{equation}
\epsilon = _E K_0^{~0} = 0~\longrightarrow E=0.
\end{equation}

So, the physical situation in this case is {\it qualitatively} and
{\it quantitatively} different than in the case of a Minkowskian
spacetime endowed with an inertial coordinates $(t,x,y,z)$.
Namely, in Minkowskian spacetime covered by an inertial
coordinates $(t,x,y,z)$ the canonical energy-momentum complex $_E K_i^{~k}$
(and other energy-momentum complexes also) {\it identically
vanishes} and for {\it any observers} we have $\epsilon = 0, ~P^i = 0.$

Thus, by using double index energy-momentum complexes, one {\it cannot
assert} that the flat Friedman universes are {\it energetic nonentity},
like a Minkowskian spacetime. All depends in the case on family of
the used observers.
\item An open dust Friedman universe.

In this case all the components of the canonical energy-momentum
complex $_E K_i^{~k}$ are different from zero    in the
conformally flat coordinates $(\tau,x,y,z)$. So, an open dust
Friedman universe surely is not an {\it energetic nonentity}.

If one calculates the ``total energy density'' $\epsilon = _E
K_i^{~k}{}u^iu_k$, matter and gravitation, for family of the
observers which are at rest in the conformally flat coordinates
$(\tau,x,y,z)$, i.e., for observers which have their 4-velocities
of the form $u^i = {\delta^i_0\over\Omega}$ in these coordinates,
then one gets
\begin{equation}
\epsilon = _E K_0^{~0} = (-){3\over 2}\alpha
a_0^2{\bigl(2\sqrt{\tau^2-r^2} - a_0\bigr)^2\over(\tau^2
-r^2)}\biggl[{r^2\over(\tau^2-r^2)^3}-{\tau^2(a_0
-2\sqrt{\tau^2-r^2})\over a_0(\tau^2-r^2)^3}\biggr].
\end{equation}
This expression is {\it negative-definite} and the formal integral
\begin{equation}
E = \int\limits_{\tau^2 - r^2 =B^2}{ _E K_0^{~0} d^3S}
\end{equation}
over hypersurface $\tau^2 - r^2 = B^2, ~B:={a_0\over 2}e^{\eta_0}>{a_0\over 2}$
is divergent to minus infinity \footnote{The hypersurface $\tau^2 - r^2 = B^2$
is an image in the conformally flat coordinates $(\tau,x,y,z)$ of
the spatial slice $\eta=\eta_0$ of the Friedman universe in the
initial coordinates $(\eta,\chi,\theta,\varphi)$.}.

The integral (53) has mathematical meaning  of the {\it global energy},
matter and gravitation, contained in the hypersurface $\tau^2 - r^2 = B^2,~B>{a_0\over 2}$
 [for observers which are at rest in the conformally flat
coordinates $(\tau,x,y,x)$ in which the line element $ds^2$ is
given by (25)].
\item A closed dust Friedman universe.

In this case also all the components of the canonical
energy-momentum complex $_E K_i^{~k}$ are different from zero
in the conformally flat coordinates $(\tau,x,y,z)$. Thus, this
universe, like an open Friedman universe, has {\it non-zero}
``energy density'' for an arbitrary set of observers, i.e., a
closed dust Friedman universe {\it is not an energetic nonentity}.

Concerning global energy of a closed dust Friedman universe we
must remember that this notion has only some {\it mathematical
meaning}, and that {\it the conformally flat coordinates} $(\tau,x,y,z)$ {\it cover entirely
only one distinguished slice} $\eta= 0$ of a closed dust Friedman
universe.

In conformally flat coordinates $(\tau,x,y,z)$ this slice is given
by
\begin{equation}
\tau=0,~(-)\infty<x<\infty,~(-)\infty<y<\infty,~(-)\infty<z<\infty.
\end{equation}
At the moment $\tau = 0$ the fundamental observers which were
at rest in the initial coordinates $(\eta,\chi,\theta,\varphi)$
are also at rest in the conformally flat coordinates
$(\tau,x,y,z)$. It is easily seen from the formulas (19)-(20) of
the Section 2A. So, for these observers the component $_E K_0^{~0}(\tau =0,x,y,z)$
represents {\it total ``energy density''}, matter and gravitation,
at the moment $\tau = 0$.

By a simple calculation one can easily get that this component $_E K_0^{~0}(\tau=0,x,y,z)$
reads
\begin{equation}
_E K_0^{~0} = {(-)384\alpha a_0^2(r^2-1)\over(r^2+1)^4}.
\end{equation}
Formal calculation of the energy contained inside of the
distinguished slice $\tau =0$ in the conformally flat coordinates $(\tau,x,y,z)$
gives
\begin{eqnarray}
E& = &\int\limits_{\tau = 0}{} _E K_0^{~0}dxdydz = (-) 1536\pi\alpha
a_0^2\int_{0}^{A}{(r^4-r^2)\over(r^2+1)^4}dr\nonumber \\
&=& {512\pi\alpha a_0^2 A^3\over(1+A^2)^3} >0.
\end{eqnarray}

$A$ can be {\it arbitrary big} but it always should be finite
because $A\longrightarrow\infty$ leads us to $\Omega\longrightarrow
0$, i.e., the limiting process $A\longrightarrow\infty$ leads to a
singularity.

Despite that, if we take the formal limit
$A\longrightarrow\infty$, then we will get $E= 0$.

But one {\it cannot conclude from this result that the closed dust
Friedman universe really has zero net global energy}.

The reasons are the following. At first, {\it one cannot
calculate} analogous global integral over any other spatial slice
 $\eta= \eta_0 = const\not=0,~(-)\pi<\eta_0<\pi$
of the closed dust Friedman universe because other slices {\it are
not entirely covered} by the conformally flat coordinates
$(\tau,x,y,z)$. We have already mentioned about this important
fact in Section 2A. Secondly, {\it we have no global conservation laws
in the domain of the relativistic cosmology}.

Thirdly, if we use an other set of observers, e.g.,
the set of observers which have their 4-velocities at the moment $\tau = 0$
\begin{equation}
u^0 = {1\over\Omega\sqrt{1-v^2}}, ~u^1 =
{v\over\Omega\sqrt{1-v^2}},~u^2 = u^3 =0,
\end{equation}
where $v = \sqrt{({dx\over d\tau})^2}$, then we will obtain for such
observers (For simplicity we will put $v = const>0$)
\begin{equation}
\epsilon = _E K_i^{~k}u^iu_k = (-){384\alpha
a_0^2\over(1-v^2)}\biggl[{r^2(1-v^2)+2v^2x^2-1\over(r^2+1)^4}\biggr].
\end{equation}

It follows from the above expression that for these observers the
``global energy'' $E$ contained in the subspace $\tau =0$ reads
\begin{eqnarray}
E &=& (-) {384\alpha
a_0^2\over(1-v^2)}\int_{0}^{\infty}\int_{0}^{\pi}\int_{0}^{2\pi}{[(1-v^2)r^2
+2v^2x^2-1]\over(1+r^2)^4}r^2\sin\theta drd\theta
d\varphi\nonumber\\
&=& {16\pi^2\alpha a_0^2 v^2\over(1-v^2)}>0,
\end{eqnarray}
i.e., it is {\it positive-definite} even for infinitesimally small
$v$.

Thus, the ``global energetic content'' in the subspace $\tau = 0$
{\it depends on the used set of the observers} which are studying
the closed dust Friedman universe.

Once more we met a situation which is {\it qualitatively and
quantitatively different} than the situation in Minkowskian
spacetime endowed with an inertial coordinates $(t,x,y,z)$.
\end{enumerate}
\vspace{0.3cm}
\section{Conclusion}

Our conclusion is that the Friedman universes {\it are not
energetic nonentity} even if we analyse these universes only with
the help of a double index energy-momentum complex. Because these universes
{\it are not asymptotically flat}, such analysis
should be performed in the {\it geometrically and physically distinguished
conformally flat coordinates} $(\tau,x,y,z)$.

We hope that we have convincingly justified this conclusion in
this paper.

The our conclusion is in full agreement with our previous
analysis of the Friedman (and also more general) universes with
the help of the {\it averaged relative energy-momentum
tensors} \cite{Gar07}.

Of course, our conclusion contradicts  the recently very popular
opinion that the {\it Friedman universes are energy-free}. Such opinion
originated from {\it incomplete analysis}
of these universes performed in the Cartesian comoving coordinates $(t,x,y,z)$ in
which only the spatial part of the {\bf FLRW} line element is
conformally flat.

By {\it incomplete analysis} we mean  the fact {\it of using only the
comoving observers} to analyse the energetic content of the
Friedman (and also more general) universes. As we have seen, using an other set of the
observers gives other, non-null local and global results for flat
Friedman universes and non-nul global results for a closed
Friedman universe.

In fact, only by using the non-comoving observers  one is able
to show that the flat Friedman universes {\it are not energetic
nonentity neither locally nor globally} and that the closed
Friedman universes {\it are not global energetic nonentity}.

Limitation to the comoving observers only {\it is not justified
physically}, e.g., an Earth's observer {\it is not a comoving
observer} in the real Universe.

We think that the conformally flat coordinates $(\tau,x,y,z)$ {\it have
much more profound  geometrical and physical meaning} than the
Cartesian comoving coordinates. Thus, in order to correctly
analyse the energy and momentum of the Friedman universes with the
help of a coordinate-dependent energy-momentum complex one should
work in these coordinates. We have done this in the present paper
for the energy.

We hope that this paper and the our previous paper \cite{Gar07}
will finish discussion about energetic content of the Friedman
universes.

\acknowledgments

This paper was partially supported by Polish Ministry of Science and
Higher Education Grant No  1P03B 04329 (years 2005-2007).

\end{document}